\documentclass[12pt]{article}
\usepackage{graphicx}
\newcommand{\beq}{\begin{equation}}
\newcommand{\eeq}{\end{equation}}
\newcommand{\bea}{\begin{eqnarray}}
\newcommand{\eea}{\end{eqnarray}}

\begin{document}
\begin{titlepage}
\begin{flushleft}
       \hfill                      {\tt hep-th/0506168}\\
       \hfill                       FIT HE - 05-03 \\
\end{flushleft}
\vspace*{5mm}
\begin{center}
{\bf\LARGE Flavor meson localization \\
\vspace{3mm}
in 5d brane-world}

\vspace*{5mm}
\vspace*{12mm}
{\large Kazuo Ghoroku\footnote[2]{\tt gouroku@dontaku.fit.ac.jp}}
\vspace*{2mm}

\vspace*{2mm}

\vspace*{4mm}
{\large ${}^{\dagger}$Fukuoka Institute of Technology, Wajiro, 
Higashi-ku}\\
{\large Fukuoka 811-0295, Japan\\}
\vspace*{4mm}

\end{center}

\begin{abstract}
We propose a brane-world, which contains flavor quarks and mesons, 
by embedding dimensionally reduced D7-brane in both the supersymmetric and 
non-supersymmetric 5d background which are obtained
as the solutions of type IIB supergravity compactified on Ad$S_5\times S^5$.
In the supersymmetric case, the RS brane can be put at any point of the fifth 
coordinate, but it is pushed to the Ad$S_5$ boundary in the non-supersymmetric
case.
We study the localization of the flavor mesons, the fluctuation-modes of 
D7-brane, on the Randall-Sundrum brane 
in these backgrounds. 

\end{abstract}
\end{titlepage}

\section{Introduction}

It would be very powerful to use the gravity/gauge
correspondence 
in getting non-perturbative insights of 
various gauge theories.
This idea has been extensively studied in 
5d gauged supergravity \cite{gppz1}-\cite{ACEP} 
which can be formulated as the superstring theory compactified on 
Ad$S_5\times S^5$. 
This theory is useful since it contains various scalar fields 
which correspond to the relevant and marginal operators of CFT.

\vspace{.3cm}
The advantage of the 5d 
supergravity is that it can be reconciled with
the brane-world proposed by Randall and Sundrum (RS) \cite{RS2}. 
In this brane-world, it is possible to consider 4d gravity
since the graviton is localized on the brane. 
In order to obtain more realistic braneworld, however,
it is desirable to set the braneworld in a background 
which corresponds, in a holographic sense, 
to the confining gauge theory with flavor quarks as in QCD. 
In such a brane-world, we expect to find mesons, 
the quark-bound states, 
which are localized on the RS brane as well as the graviton.

\vspace{.3cm}
Recently, a new idea to introduce flavor quarks has been proposed by
Karch and Katz \cite{KK}
by embedding a D7-brane 
as a probe which is wrapping a contractible $S^3$ 
of $S^5$ in the AdS$_5\times S^5$ background in order to avoid
the RR tadpole problem. At the price of this embedding, 
a tachyonic mode
appears on the D7-brane, but the system is stable since the mass
is within the Breitenlohner-Freedman bound \cite{BF} of the AdS background.
Meanwhile, the brane-world is set in the AdS$_5$ accompanied with the
compact $S^5$ as the inner space. 
In a similar sense, the $S^3$ is considered as
the inner space of the embedded D7-brane, and the action of the D7-brane
can be reduced to 5d by integrating over the coordinates of $S^3$.
Then we can set a system of supergravity and D7-brane which are
dimensionally reduced to five dimensions.

\vspace{.2cm}
The purpose of the present paper is to study the problem of embedding
the reduced D7-brane in the 5d background of an appropriate brane-world
according to the procedure given in \cite{chiral}. And we also
study the localization of the flavor mesons, which are
given as fluctuation modes of the D7-brane, on the RS brane.
The analysis is performed for the simple background
studied before in \cite{chiral}.

\vspace{.2cm}
In Section 2, we set up our brane-world model. 
In Section 3, in this brane-world, the dimensionally reduced 
D7 brane is embedded. In Section 4,
the localization of the mesons are studied, and summary is given
in the final section.

\section{Setting of a brane-world}
Here we set up our model of a simple brane-world to embed a
dimensionally reduced D7-brane.
Consider the 10d type IIB supergravity which includes
the dilaton ($\Phi$), axion ($\alpha$) and five form field strength
to obtain ${\cal M}_5\times S^5$ space-time with the following
form of the Einstein frame metric,
\beq 
ds^2_{10}= 
{r^2 \over L^2}A^2(r)\eta_{\mu\nu}dx^\mu dx^\nu +
\frac{L^2}{r^2}\left( dr^2+r^2 d\Omega_5^2\right)  \ ,
\label{non-SUSY-sol}
\eeq 
where $A(r)$ denotes a deviation of ${\cal M}_5$ from the exact Ad$S_5$.
The ${\cal M}_5$ part, $ds_5^2={r^2 \over L^2}A^2(r)
\eta_{\mu\nu}dx^\mu dx^\nu + \frac{L^2}{r^2}dr^2$, is rewritten by
using the new variable $y$, defined as $r=e^{-y/L}$, as follows
\beq
 ds^2= \tilde{A}^2(y)\eta_{\mu\nu}dx^{\mu}dx^{\nu}
           +dy^2  \, , \label{metrica}
\eeq
where $\tilde{A}^2={r^2 \over L^2}A^2(r)$. 
This metric is also obtained from the following 5d theory,
\beq
   S_{\rm g}=\int d^{4}x^{\mu}dy~\sqrt{-g}
   \left\{{1\over 2\kappa_5^2}R  
    -{1\over 2}(\partial\Phi)^2-{1\over 2}e^{2\Phi}(\partial\alpha)^2-V 
\right\} , 
\label{acg}
\eeq
where $V=-6/(L^2\kappa_5^2)$.\footnote{
In the following, we set as $V=-3$ for $\kappa_5^2=2$
and $L=1$.} This action is obtained from 10d supergravity
by integrating out the compact $S^5$ part. And this is
also equivalent to the 5d gauged
supergravity written by dilaton and axion only. 
In this case, the potential $V$ in $S_{\rm g}$ is generally given as 
\beq
 V(\phi)={v^2\over 8}\sum_i\left({\partial W\over \partial \phi_i}\right)^2
   -{v^2\over 3}W^2 , \label{superPot}
\eeq
where $W$ represents the superpotential of the theory
and 
$v$ denotes the gauge 
coupling~\footnote{The gauge coupling parameter $v$ is fixed 
from the AdS$_5$ vacuum \cite{FGPW} 
by fixing the radius of AdS as a unit length. Here we set as $v=-2$.}.
In the present case, $\phi_i=\{\Phi,~\alpha\}$ and $V=-3$.

\vspace{.2cm}
The supersymmetric solutions of $S_{\rm g}$
are given by solving the following first order equations \cite{ST},
\beq
 \phi_i'={v\over 2}{\partial W\over\partial\phi_i}, \qquad
    {\tilde{A}'\over \tilde{A}}=-{v\over 3}W,  \label{first-order0}
\eeq
where $'=d/dy$. 
However, for some non-supersymmetric
solutions, there exists $W$ which satisfies the above relations
(\ref{superPot}) and (\ref{first-order0}). This is the case
known as the fake supersymmetry \cite{FNSS}. 

\vspace{.2cm}
The solutions are obtained under the ansatzs, $\phi_i=\phi_i(y)$.
Here we consider both the fake and the true supersymmetric cases.
The AdS$_5$ solution is considered as a simple example
for the supersymmetric case,
\beq
A=1,~\Phi=\Phi_0,~ \alpha=0,~~ \label{A}
\eeq
where $\Phi_0$ is a constant and $W=-3/2.$ 
As for the fake-supersymmetric case, we adopt the following solution
\cite{FNSS,chiral},
\beq
 A=\left(1-({r_0\over r})^8\right)^{1/4},~~
 e^{\Phi}=\left({1+({r_0/r})^4\over 1-({r_0/r})^4}\right)
    ^{\sqrt{3/2}},~~\alpha=0, \label{C}
\eeq
where $r_0$ is a constant and 
\beq
 W=-{3\over 2}\cosh(\sqrt{{8\over 3}}\Phi). \label{w-nonsusy}
\eeq

\vspace{.2cm}
The solutions of this 5d theory can be easily
lifted up to 10d. However, when the other scalars are
added and they take some non-trivial
configurations, it is non-trivial to 
lift up the solutions to 10d, and $S^5$ has been deformed to a 
complicated geometry \cite{PW}. In this case, it is not simple to embed 
the D7 brane in this 10d background.

\vspace{.2cm}
The brane-world solution is obtained by solving the following system,
\beq
S=S_{\rm g}+S_{\rm b},
\eeq
where $S_{\rm b}$ is the action of the RS-brane.
Here, we set $S_{\rm b}$ as follows,
\beq
    S_b = -{v}\int d^4x dy\sqrt{-g}W(\phi_i)\delta(y-y_b). \label{baction}
\eeq
where $y_b$ denotes the brane position. In solving the equations of motion
of $S_g+S_b$, the role of the brane action is to give the
boundary conditions for the bulk solution of $S_g$. 
Meanwhile, these conditions 
for $\Phi$ and $A$ are equivalent to the above equations 
(\ref{first-order0}) evaluated just on the point $y=y_b$. 
Then, at any point of $y_b$, 
all the solutions of (\ref{first-order0}) satisfy
the boundary conditions given by $S_b$ of the form of (\ref{baction})
\cite{G1}. So it is enough to solve equations (\ref{first-order0})
to obtain the brane-world solution which satisfies the boundary
conditions at $y_b$. In other words, it is possible to put the brane 
at any point on the $y-$axis. This situation is not changed even if
the "superpotential" $W$ is given by (\ref{w-nonsusy}) with the 
fake-supersymmetic solution given above.

\vspace{.3cm}
\section{Reduced D7-brane and its embedding}
In \cite{chiral}, we have given an explicit embedding of D7-brane as a probe
in the 10d background (\ref{non-SUSY-sol}).
Rewrite the metric (\ref{non-SUSY-sol}) as
$$
ds^2_{10}= 
{r^2 \over L^2}A^2(r)\eta_{\mu\nu}dx^\mu dx^\nu +
\frac{L^2}{r^2}\left( dr^2+r^2 d\Omega_5^2\right)  \ 
\label{non-SUSY-sol-2}
$$%
\beq 
= 
{r^2 \over L^2}A^2(r)\eta_{\mu\nu}dx^\mu dx^\nu +
\frac{L^2}{r^2}\left( \sum_{m=4}^{7} (dX^m)^2+\sum_{i=8}^{9} (dX^i)^2\right), 
\eeq
and the four dimensional coordinates $\{X^m\}$
are further rewritten in the polar coordinate as 
\beq
\sum_{m=4}^{7} (dX^m)^2=d\rho^2+\rho^2d\Omega_3^2, 
\eeq 
where $\rho^2=\sum_{m=4}^7(X^m)^2$. Then, the world-volume of the D7-brane is 
taken as $\{x^{\mu}, \rho, \Omega_3\}$.
As mentioned above, the D7-brane is wrapping on
the contractible $S^3$, and 
we notive the relation $r^2=\rho^2+(X^8)^2+(X^9)^2$ in this case.

\vspace{.3cm}
The remaining part of the embedding procedure is to determine the
shape of the D7-brane in its
outer two dimensional space, $(X^7,X^8)$. This is performed by solving
the equations of motion of two scalar fields 
in the D7-brane action, which
is written as \cite{chiral}
\beq
S_{\rm D7}= -\tau_7 \int d^8\xi \left(e^{-\Phi} \sqrt{\cal G} 
      +{1\over 8!}\epsilon^{i_1\cdots i_8}A_{i_1\cdots i_8}\right) \ ,
\label{D7-action}
\eeq
where ${\cal G}=-{\rm det}({\cal G}_{i,j})$, $i,~j=0\sim 7$. ${\cal G}_{i,j}$ 
denotes the induced metric 
and $\tau_7$ is 
the tension of D7 brane. 
The eight form potential $A_{i_1\cdots i_8}$, 
which is Hodge dual to the axion, couples to the 
D7 brane minimally. It is obtained 
as $F_{(9)}=dA_{(8)}$ in terms of the Hodge dual field strength $F_{(9)}$ 
\cite{GGP}.
We take the gauge, $\xi^i=\{~x^{\mu},~ \rho,~ \theta^j~\}$, 
where $\theta^j ~(j=1\sim 3)$ denote the angle variables of $S^3$,
and we make an ansatz for the two scalar fields $X^8(\xi)$ and $X^9(\xi)$ as
$X^9\equiv w(\rho)$ and $X^8=0$, without
loss of generality. 
Then, the induced 
metric and the explicit form of the action (\ref{D7-action}) are obtained as
\cite{chiral}  
\beq
ds^2_8=e^{\Phi/2}
\left(
{r^2 \over L^2}A(r)^2~\eta_{\mu\nu}dx^{\mu} dx^{\nu} + 
\frac{1+(w')^2}{r^2}L^2 d\rho^2 
+\frac{\rho^2}{r^2}L^2 d\Omega_3^2 \ .
 \right)
\eeq
\beq
S_{\rm D7} =-\tau_7~\int d^8\xi~\sqrt{\epsilon_3} \left(L_{D7}^{\rm cl}
                        +L_{D7}^{\rm fluc}\right), 
\eeq
\beq 
\quad
L_{D7}^{\rm cl}=\rho^3
\left( e^{\Phi}~A(r)^4 \sqrt{ 1 + (w')^2 }+C_8(r)\right) 
\label{SUSY-LD7}
\eeq
where $C_8(r)$ denotes the contribution of the eight form potential, and
the fluctuations of D7 around its classical configuration are
included in $L_{D7}^{\rm fluc}$.

\begin{table}
\begin{center}
\renewcommand{\arraystretch}{1.2}
\setlength{\arrayrulewidth}{.3\tabcolsep} 
\begin{tabular}{*{10}{c}}\hline
        0&1&2&3&4&5&6&7&8&9\\
        \hline 
        \multicolumn{4}{c|}{RS-brane}&&&&&&\\ \hline
        \multicolumn{4}{c}{$x^\mu$}&
        \multicolumn{6}{|c}{$r$, $\Omega_5$}\\ \hline
        \multicolumn{8}{c|}{D7-brane}&&\\ \hline
        \multicolumn{4}{c}{$x^\mu$}&
                \multicolumn{4}{|c|}{ $\rho, \Omega_3$ } &
                \multicolumn{2}{c}{$X^i$}\\ \hline
\end{tabular}
\caption{\label{tab:indices}Coordinate
conventions}
\end{center}
\end{table}

\vspace{.3cm}
Here 
$w(\rho)$ 
is solved independently of the angle variables $\theta^i$ on $S^3$, so
the form of the solution $w$ is preserved even if
the action was reduced to 5d by integrating over
$\theta^i$. 
As a result, we can reduce the above action as
\beq
S_{\rm D7}^{(5d)} =-2{\pi^2}~\tau_7\int d^4x^{\mu}d\rho~
      \left(L_{D7}^{\rm cl}+L_{D7}^{\rm fluc}\right). \label{SUSY-SD5}
\eeq
Thus we obtain 
the D7-brane action which is dimensionally reduced to the 5d. Here
we notice the relation, $r^2=\rho^2+w^2(\rho)$, for $r$ and $\rho$.
We preserve the interpretation of $r$ or $\rho$ 
as the energy scale
of the field theory on the brane, so
we select the solution of $w(\rho)$ such that it could give the one to one
correspondence between $r$ and $\rho$ \cite{chiral}.

\vspace{.3cm}
We are now ready to embed the reduced D7-brane in the 5d brane-world.
As done in 10d, it is embedded 
as a probe such that the given background configuration is not altered. 
This is performed by solving
the equation of motion for $w$, which is 
obtained from $S_{D7}$ only in the brane-world background. 
However we impose $Z_2$ symmetry with respect to the coordinate $y$ as,
\beq
   w(y)=w(|y-y_b|).
\eeq
Then the equation of $w$ is obtained as
\beq
 \tilde{\tau}_7\left(\delta_w L_{D7}^{\rm cl}
  -F(\rho){2 w'G(\rho)\over (1+w'~^2)^{3/2}}\delta(\rho-\rho_b)\right)=0 , 
\label{w-equation2}
\eeq
where $w'=dw/d\rho$,
$\delta_w L_{D7}^{\rm cl}=\partial_w L_{D7}^{\rm cl}-\partial_{\rho}
         \left(\partial_{w'}L_{D7}^{\rm cl}\right)$,
$F(\rho)=\rho^3 e^{\Phi} A^4$ and $ G(\rho)=(\rho+ww')^2/\rho^2$. And 
$\rho_b$ denotes the position of the brane on the coordinate $\rho$,
defined as $e^{-2y_b/L}(=r_b^2)=\rho_b^2+w^2(\rho_b)$. 
The $\delta$-function of Eq.(\ref{w-equation2})
comes from the $Z_2$ symmetry at
$y=y_b$, and the factor $G(\rho)$ appears when the 
$\rho$-derivative is changed to the $y$-derivative. 
For $\rho<\rho_b$,
the field equation for $w$,
$\delta_w L_{D7}^{\rm cl}=0$, is written as
\beq
  w''+(1+w'~^2)\left[{3\over\rho}w'+2K_{(1)}(\rho w'-w)
\right]=0
\label{w-equation3}
\eeq
where $K_{(1)}=\partial_{r^2}\log(e^{\Phi}A^4)$, and the boundary condition, 
\beq
w'(\rho_b)=0  \ . \label{condi-2}
\eeq

\vspace{.2cm}
For the super-symmetric background (\ref{A}), $K_{(1)}=0$. Then we find the
solution $w'(\rho)=0~ (w=$const.) which preserves the super-symmetry of 
the system \cite{chiral}, and
the condition (\ref{condi-2}) is satisfied at any $\rho$. 
While, for the non-supersymmetric background 
(\ref{C}), $K_{(1)}\neq 0$, then $w'(\rho)\neq 0$ and the boundary condition
is not satisfied except at $\rho=0$ and $\rho=\infty$ since $w(\rho)$
decreases monotonically with $\rho$. 
For finite $\rho_b$, 
then, the condition (\ref{condi-2}) selects the super-symmetric solution.
And the RS brane in the non-supersymmetric background is pushed
toward the boundary, $\rho_b=\infty$.

\vspace{.3cm}
As another possible boundary condition, we may consider 
$G(\rho_b)=0$, and we get
\beq
 w(\rho_b)w'(\rho_b)+\rho_b=0. \label{condi-3}
\eeq
This is equivalent to $\partial_{\rho}r(\rho_b)=0$.
Meanwhile $r(\rho)$ must increase monotonically with $\rho$ since it should
be a single valued function of $\rho$ due to the requirement of the
one to one correspondence between $r^2(=\rho^2+w(\rho)^2)$ and $\rho$. Thus,
the points which satisfy the condition, $\partial_{\rho}r(\rho_b)=0$, 
are again
restricted to $\rho_b=0$ and $\rho_b=\infty$. Then this condition 
is not useful as the boundary condition. By the same reason, $F(\rho_b)=0$
is also rejected as a useful boundary condition. After all, 
we find that only the supersymmetric solution satisfies the 
meaningful boundary 
condition (\ref{condi-2}).

\vspace{.2cm} 
Here we remember the asymptotic form of $w$ at large $\rho$, 
\beq
  w(\rho)=m_q+c/\rho^2, \quad c=-\langle\bar{\Psi}\Psi\rangle
\eeq
and the meaning of the parameters in it. Namely $m_q$ represents the quark-mass
and $c$ is the vacuum expectation value of bi-linear operators of 
quark fields ($\Psi$) 
in the dual gauge theory. Then we can see
$\langle\bar{\Psi}\Psi\rangle=0$ and $w=m_q$ in the supersymmetric case.
As a result, the chiral symmetry is preserved.

\vspace{.3cm}
In the above analysis, we have ignored the interaction between D7 and
RS brane since D7 should be treated as a probe. However, the situation
is not changed even if we solve the equation of $w$ by taking into account of 
the brane action as,
$S_{D7}+S_{b}~$, since $S_{b}$ does not depend on $w$. In other words,
the shape of the D7-brane is independent of the RS brane.
So we need the back-reaction from the bulk in order to change the
situation of the embedding. But this is out of the present work.


\section{Meson localization}
In the next,
we study the fluctuation-modes of the embedded
D7 brane. Some of them are trapped on the RS brane, and they might be
observed as mesons in our 4d world. Then, the trapped modes
are defined as the normalizable one for the integration over $\rho$.

\subsection{For finite $\rho_b$} 
In this case, only the supersymmetric embeddings are allowed. So the 
background is given by (\ref{A}), and 
the quadratic parts of the fluctuations, $\phi_8=X^8,~\phi_9=X^9-w$
and vector, are written as
\beq
  L_{D7}^{(2)}=\rho^3 \left\{{1\over 2}e^{\Phi}{L^2\over r^2}
   \sum_{i=8,9}(\partial\phi_i)^2+F_{ab}F^{ab}+\cdots\right\}  
\label{quadra-action}
\eeq
where $w$ is a constant, $w(\rho)=w(\infty)\equiv m(=m_q/2\pi)$.
The dots denote the higher order
terms which can be neglected here. 

\vspace{.3cm}
For the scalar, we obtain the same field equation for $\phi^8$ and
$\phi^9$, then they are denoted by $\phi$ for the simplicity. The field-equation 
is written as \cite{KMMW}
\beq
 \partial_{\rho}^2\phi+
{3\over\rho}\partial_{\rho}\phi
 =-{M^2R^4\over({m}^2+\rho^2)^2}\phi \, , \label{fl-scalar}
\eeq
where $M^2$ is defined as 
$\eta^{\mu\nu}\partial_{\mu}\partial_{\nu}\phi=M^2\phi$. 
In the present case, $\phi$ is $Z_2$ symmetric at the brane position
$y_b$, so
we must solve the equation (\ref{fl-scalar})
by imposing the following boundary condition, 
\beq
\phi'(\rho_b)=0. \label{bound-f}
\eeq

Before studying the trapping in the brane-world by using the 
boundary condition (\ref{bound-f}), we consider the normalizable modes
which could be observed at the boundary $\rho_b=\infty$. This analysis
is useful for finding the localized modes on the brane since the 
mass eigen-values of the
localized modes on the branes at different $\rho_b$ should be related
to each other by the smooth functions of $\rho_b$ 
in all the region $0<\rho_b<\infty$.
In this sense, the spectra, which should be observed on the boundary,
are considered as the one at the starting point.

The Eq.(\ref{fl-scalar}) is solved as
\beq
 \phi=(\rho^2+{m}^2)^{-\alpha}\left(c_1F(-\alpha,-\alpha+1,2;-\rho^2/{m}^2)+c_2\rho^{-2}F(-\alpha-1,-\alpha,0;-\rho^2/{m}^2)\right),
\label{sol-q-0}
\eeq
where $c_1$ and $c_2$ are arbitrary constants,
and $\alpha=(1+\sqrt{1+M^2L^4/{m}^2})/2$. In order to find the trapped
modes, consider the normalizability condition for $\phi$,
\beq
 \int_0^{\rho_b}d\rho~ \rho^3 
       \left({L^2\over \rho^2+{m}^2}\right)^2\phi^2(\rho) < \infty .
\label{normalization-c}
\eeq
We estimate this condition for (i) $m\neq 0$ and (ii) $m=0$ separately.
For the case of (i), the above integral near $\rho\sim 0$
is approximately estimated as
\beq
 \int_0 d\rho~ \rho^3 
\phi^2(\rho) < \infty . \label{normalization-c0}
\eeq
Meanwhile, the solution (\ref{sol-q-0}) is expanded near $\rho\sim 0$ as, 
\beq
\phi= c_1(1+c_1^1\rho^2+\cdots)+c_2\rho^{-2}(1+c_2^1\rho^2+\cdots) ,
   \label{small}
\eeq
where $c_1^1, c_2^1$ are the calculable coefficients.
Then we find that the solution of $c_2=0$ satisfies (\ref{normalization-c0}). 
However, for the case of (ii), the condition (\ref{normalization-c0}) is replaced by 
\beq
 \int_0 {d\rho\over \rho}~ 
\phi^2(\rho) < \infty . \label{normalization-c0-2}
\eeq
Then we must take as $c_1=c_2=0$, namely $\phi=0$, 
to satisfy (\ref{normalization-c0-2}). 
In other words, there is no
localized state in the case of $m=0$ for the supersymmetric case.
So the interesting case is restricted to the case of massive quark.

\vspace{.3cm}
On the other hand,
at large $\rho$, the normalizability condition 
(\ref{normalization-c}) is approximated for any $m$ as,
\beq
 \int ^{\infty}{d\rho\over \rho}
       \phi^2(\rho) < \infty . \label{normalization-c2}
\eeq
And the solution of $c_2=0$,
the first term of (\ref{sol-q-0}), is expanded 
at large $\rho$ as, 
\beq
 \phi/c_1=b_0(\alpha)(1+b_1(\alpha)/\rho^2+\cdots)+d_0(\alpha)
\rho^{-2}(1+d_1(\alpha)/\rho^2+\cdots),
 \label{asymptotic}
\eeq
where the coefficients $b_0(\alpha)$ and 
$d_0(\alpha)$ are the functions of $\alpha$.
From (\ref{normalization-c2}), we demand $b_0(\alpha)=0$. As a result,
we get $\alpha=n+1$ with the integer n \cite{KMMW}, and we 
thus find infinite series of discrete meson mass.

\begin{figure}[htbp]
\begin{center}
\voffset=15cm
  \includegraphics[width=8cm,height=5cm]{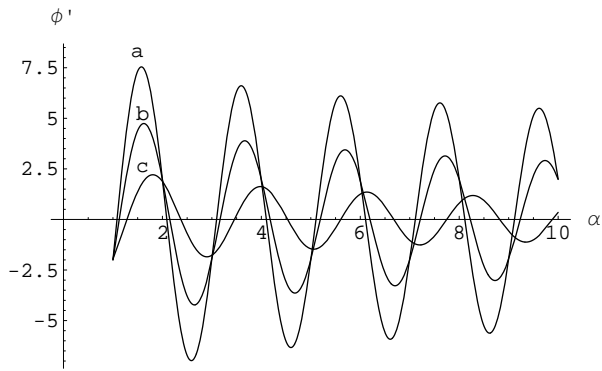} 
\caption{ $\phi'$ vs $\alpha$ 
for $m=1$. The curves a, b and c represent $\left\{\phi'\times 10^9,
\rho_b=10^3\right\}$, $\left\{\phi'\times 10^6, \rho_b=10^2\right\}$ and 
$\left\{\phi'\times 10^3, \rho_b=10\right\}$ respectively.
\label{mass-spectra} }
\end{center}
\end{figure}

\vspace{.3cm}
We now return to the case of the brane-world. 
In this case, $\rho_b$ is finite and we need the boundary
condition (\ref{bound-f}) at $\rho_b$ instead of the normalizable
condition at large $\rho$ (\ref{normalization-c2}) given above. 
In order to satisfy (\ref{bound-f}), 
$b_0(\alpha)/d_0(\alpha)$ 
should be fixed to an appropriate value $f$, which depends on $\rho_b$,
\beq
  b_0(\alpha)/d_0(\alpha)=f(\rho_b),
\eeq
and $f$ should satisfy the condition, $f(\infty)=0$. From this equation,
we could find the mass eigen-values.
Here, we can see this statement by a numerical
analysis. For different three values of 
$\rho_b$, the value of $\phi'$ as a function of $\alpha$ 
is shown in the Fig.\ref{mass-spectra}. 
We find that the values of $\alpha$ at each zero point of
$\phi'$ shift to larger $\alpha$ side from the integer point, 
which are given by $\alpha=n+1$ for $\rho_b=\infty$,
smoothly when $\rho_b$ decreases.

We could conclude as follows.
Also on the RS brane, in the present supersymmetric braneworld, 
we can see the meson mass-spectra,
which shift to the massive side
when the brane moves to the smaller $\rho_b$.

\subsection{Non-supersymmetric background}
For non-supersymmetric background, we must take $\rho_b=\infty$.
In this case, the field equations for $\phi^8$ 
and $\phi^9$ are complicated, but the equations for both fields
have the same asymptotic solution with the one given 
for the supersymmetric case
by the equation (\ref{small}) at small $\rho$
and (\ref{asymptotic}) at large $\rho$ respectively. 
In the present case, however, the normalizable
condition has a different form from (\ref{normalization-c}) and it is
given as
\beq
 \int_0^{\infty}d\rho \rho^3 e^{\Phi}\left({L^2\over Ar^2}\right)^2
       \phi(\rho)^2 <\infty  \label{bc-nonsusy}
\eeq
where $\phi$ denote $\phi^8$ or $\phi^9$. 

\vspace{.3cm}
In the small $\rho$ region, 
we find the following approximated condition
\beq
   \int_0 d\rho \rho^3 \phi(\rho)^2 <\infty \ \ . \label{small-2}
\eeq
It should be noticed that this condition is not changed even if $m=0$
because, in the limit of $\rho\to 0$, we find
$$r^2=\rho^2+w^2(\rho)\to w^2(0)\neq 0$$
for the non-supersymmetric case and that $e^{\Phi}$ and
$A$ approach to the constant \cite{chiral}. 
This is an important point
to be discriminated from the supersymmetric case. And (\ref{small-2})
is satisfied by the first one of the two independent
solutions given in (\ref{small}).
Then we expect to find normalizable modes or the meson spectra 
on the RS brane for any value of $m$.

\vspace{.2cm}
At large $\rho$, (\ref{bc-nonsusy})
is approximated by (\ref{normalization-c2}),
and the asymptotic solution for $\phi$ is also the same form with
(\ref{asymptotic}). Then, we can choose the second series of 
(\ref{asymptotic}) as the normalizable solution, and
many number of normalizable modes are found
for both $\phi^8$ and $\phi^9$. Especially, for the case of $m=0$,
we find one mass-less mode for $\phi^8$ as the Nambu-Goldstone 
boson due to the spontaneous chiral symmetry breaking. Actually, these points
have been shown in \cite{BGN,BEEGK} when the RS brane is absent. In the
present case, 
from the $Z_2$ symmetry, we need new conditions for the modes are 
$\partial_{\rho}\phi^8|_{\rho=\infty}=\partial_{\rho}\phi^9|_{\rho=\infty}=0$.
And all the normalizable solutions satisfy this.
Then it would be 
straightforward to find the meson-spectra on the RS brane.

\section{Summary}
Here, we consider a class of background solutions,
$\mathcal{M}_5\times S^5$, of IIB superstring
theory in order to construct a braneworld in which mesons are trapped
on the RS brane. The brane-world is set in the uncompactified
5d part $\mathcal{M}_5$ of these solutions. 
Its effective
action can be obtained from the 10d theory by integrating over 
the inner compact space, $S^5$.
They are also obtained from the
5d gauged supergravity with the dilaton and the axion. 
Any solution of this 5d theory can be easily lifted up to the one of 
the 10d theory on $\mathcal{M}_5\times S^5$. On the same footing,
the action of the D7-brane is reduced to 5d by integrating
over $S^3$, which is regarded as the inner compact space of the
D7-brane here.

After setting the stable RS brane, which can be put at any position
of the fifth coordinate $\rho_b$ we like, for the 5d 
background given here,
the reduced D7 brane is embedded in this background 
as a probe in the sense that the D7 brane does not
change the given 5d configuration. 
The embedded form is obtained 
by solving the scalar field equation of the
D7 brane action. 
For finite $\rho_b$, only the supersymmetric embedding is allowed and
the quark mass is arbitrary in this case. The chiral symmetry of 
the corresponding gauge theory is then preserved for the mass-less quark. 
When the quark mass is finite, the series of meson mass-spectra
are observed as trapped states on the RS brane.

\vspace{.3cm}
In the non-supersymmetric case, on the other hand, the brane is retained
on the boundary, $\rho_b=\infty$. In this case, the situation is similar
to the case of 10d gauge/gravity correspondence. We can then observe
the spontaneous chiral symmetry breaking and
infinite series of the trapped mesons on the RS brane. And we assure
the existence of 
the 
Nambu-Goldstone boson, which is generated as a result of
the spontaneous chiral symmetry breaking. 

\vspace{.3cm}
In any case, we could show a model of a braneworld which includes flavor
quarks and their bound states on the RS brane. It would be an interesting
problem to study the brane-world cosmology including
hadrons in terms of the model presented here. 
We will discuss on this point in the near future.

\vspace{.3cm}
{\bf Acknowledgments:}~
The author would like to thank M. Tachibana and M. Yahiro 
for useful discussion.
This work has been supported in part by the Grants-in-Aid for
Scientific Research (13135223)
of the Ministry of Education, Science, Sports, and Culture of Japan.


\end{document}